\documentclass[12pt]{iopart}

\expandafter\let\csname equation*\endcsname\relax
\expandafter\let\csname endequation*\endcsname\relax

\usepackage[utf8]{inputenc}
\usepackage[T1]{fontenc}
\usepackage{amsmath,amssymb,array}
\usepackage{multirow}
\usepackage{color}
\usepackage{bm}
\usepackage[pdftex]{graphicx}
\usepackage{ulem}

\newcommand{\name}[1]{\textbf{#1}}
\renewcommand{\contentsname}{\centerline{Contents}} 
\newtheorem{thm}{Theorem}
\newtheorem{defn}{Definition}

\begin{document}

\title[Quantum digital signature based on single-qubit]{Quantum digital signature based on single-qubit without a trusted third-party}

\author{Wusheng Wang}
\address{Graduate School of Mathematics, Nagoya University, Furo-cho, Chikusa-ku, Nagoya, 464-8602, Japan}
\address{School of Data Science, The Chinese University of Hong Kong, Shenzhen, Longgang District, Shenzhen, 518172, China}
\author{Masahito Hayashi}
\address{School of Data Science, The Chinese University of Hong Kong, Shenzhen, Longgang District, Shenzhen, 518172, China}
\address{International Quantum Academy, Shenzhen 518048, China}
\address{Graduate School of Mathematics, Nagoya University, Furo-cho, Chikusa-ku, Nagoya, 464-8602, Japan}
\address{Author to whom any correspondence should be addressed.}
\eads{\mailto{hmasahito@cuhk.edu.cn}}

\vspace{10pt}
\begin{indented}
\item[] July 2024.
\end{indented}

\begin{abstract}
Digital signatures are a powerful cryptographic tool widely employed across various industries for securely authenticating the identity of a signer during communication between signers and verifiers. 
While quantum digital signatures have been extensively studied, the security still depends on a trusted third-party. To address this limitation and enhance the applicability in real-world scenarios, here we propose a novel quantum digital signature protocol without a trusted third-party to further improve the security.
We note that a quantum one-way function can work appropriately in digital signature due to the intrinsic non-cloning property for quantum states. Secret keys in the protocol are constituted by classical private keys and quantum public keys because we assume that no user is trusted in the protocol.
We prove that the protocol has information-theoretical unforgeability. Moreover, it satisfies other important secure properties, including asymmetry, undeniability, and expandability.
\end{abstract}

\vspace{2pc}
\noindent{\it Keywords\/}: 
quantum digital signature,
single-qubit operation,
authentication,
information-theoretic quantum one-way function,
unforgeability

\maketitle

\section{Introduction}
Digital signatures as an important primitive in cryptography has been highly developed\cite{1}. Furthermore, it is an imperative construction
block in some more complex systems. For example, the next-generation information technology, blockchains, takes digital signature 
as a basic component\cite{2}\cite{3}\cite{4}. In a digital signature protocol, we mainly have two parties, one is the signer, and the other one is
the verifier. Then, the signer wants to make a public message. The aim of the protocol is that the verifier can judge if the 
message is really sent by the signer.

The security of classical digital signatures is usually ensured by one-way function\cite{5}, which can always be evaluated 
in one way while the inverse only succeeds probabilistically. However, due to the rapid and extensive development of
quantum computation and quantum information, many classical one-way functions are no longer safe for quantum computers. For
example, the Rivest-Shamir-Adleman(RSA) algorithm and elliptic curve algorithm, which are necessary for many primitives in classical cryptography, can
be cracked easily by the Shor algorithm\cite{6}. Scientists also endeavored to develop many quantum-resisting algorithms\cite{7}
\cite{8}\cite{9}\cite{10}, 
whereas they can not prove the non-existence of a quantum algorithm that can crack them efficiently. Hence, it is urgent to 
develop the quantum digital signature to improve the security of modern cryptography.
Experimental quantum 
digital signatures have taken a lot of steps in recent works. For example, long-distance over 100 km quantum digital signatures
were realized with photons\cite{Yin2017}\cite{Ding2020}. High signing rates were also achieved using fiber link
\cite{an2018}\cite{richter2021}. In addition, an efficient silicon chip-based quantum digital signatures network was obtained
\cite{du2024}.

It deserves to be mentioned that there are some authentication methods based on quantum physical unclonable functions
\cite{11}\cite{12}\cite{13} inspired by Haar random unitary\cite{14}. However, these methods are not typical authentication 
because messages do not exist in the protocols.
Recently, different kinds of quantum digital signature schemes sprang up. 
Zhang et al. proposed the signature based on quantum entanglement\cite{15}.
Lu et al. introduced the signature with orthogonal basis\cite{16}.
Li et al. combined the quantum digital signature with the quantum error correction\cite{17}.
Xin et al. gave a signature based on the quantum asymmetric encryption\cite{18}, which is computationally secure against quantum computers.
Yin et al. realized a quantum secure network with a quantum digital signature based on secret sharing\cite{19}.
Qin et al. designed a quantum digital signature based on a hash function and quantum key distribution \cite{20}. 
However, a trusted third-party is necessary for all the above schemes, which fairly compromises the fairness and security
for users. 
A trusted third-party would further undermine a larger system that needs digital signatures. 
We name an example of
blockchains again. Blockchains need decentralization as the key characteristic. That is, each node, or user, has the same
weight in a blockchain system. 
A trusted third-party means a more powerful user. 
Thus, in this paper, we propose a
quantum digital signature protocol without the trusted party to solve the problems mentioned above. 
To ensure the fairness of the users in our protocol, we 
employ a private key and a public key.
A one-way function plays a key role in classical digital signature, and it
maps the private key to the public key.

Although recent studies \cite{morimae2022,Morimae,MY22,CM24,KT23} studied a general framework for 
quantum digital signature based on quantum public key
their security is based on computational security based on quantum computer and it does not present a concrete protocol.
To enhance the study of quantum digital signatures, 
this paper employs information-theoretic security \cite{gottesman2001}
and presents a simple protocol.
When the adversary collects a sufficient number of copies of the quantum public key, 
the adversary can identify the form of the quantum public key
by applying the quantum tomography.
However, to make the above identification, the adversary colludes with so many players.
If the number of players colluding with the adversary 
is limited to a certain number, it is not easy to identify the public key.
For example, the papers \cite{KKNY05,kawachi} proposed a quantum public key based on permutation group, and the paper \cite{HKK08} showed
information-theoretical indistinguishability with a limited number of copies of quantum public key.
In this paper, to present our information-theoretically secure protocol, we propose a concept of information-theoretic quantum one-way function 
mapping classical information to quantum states.
In our protocol, an information-theoretic quantum one-way function
works as an alternative concept of one-way function.

A protocol is called computationally secure if an adversary cannot crack it within
polynomial time. On the other hand, a protocol is called 
information-theoretically secure if an adversary cannot crack it even given unbounded time. 
The definition of information-theoretic unforgeability for a quantum digital signature protocol is given in Section 
\ref{S2} and we show our protocol satisfies such unforgeability in Section \ref{S4}.
Our idea is to use the eigenvalues and measurement
basis as the private key, and the eigenstates on single qubit system
as the public key. 
Then, we use a part of the private key as the signature
such that one-to-one correspondences exist between messages and signatures. 
Our protocol can be implemented only with single qubit physical operations, which does not require any entangling operation.
Although it is impossible to perfectly identify 
the private key from the public key,
it is not so simple to evaluate how much the adversary knows 
the private key from the public key.
In addition, there is a possibility that 
the adversary colludes with several users having the public key.
For analysis of such a case, 
we need to study the state identification problem from several copies.
For this aim, we employ covariant state family model 
\cite[Chapter 4]{Holevo}, \cite[Chapter 4]{Group2},
and derive the optimal identification probability.
Based on this analysis, 
we clarify when an adversary cannot forge the signature of the signer except for a given small probability.

\begin{table}[t]
\caption{Comparison with existing studies}
\label{T1}
\begin{center}
\begin{tabular}{|l|c|c|c|}
\hline
 & Trusted & Message & Adversary's \\
 & third party & transmission & computation power\\
\hline
\cite{11}\cite{12}\cite{13} 
& Not needed& No & Unlimited \\
\hline
\cite{15,16,17,18,19,20}
& Needed& Yes & Unlimited \\
\hline
\cite{morimae2022,Morimae,MY22,CM24,KT23}
& Not needed& Yes & Limited \\
\hline
This paper
& Not needed& Yes & Unlimited \\
\hline
\end{tabular}
\end{center}
\end{table}

The rest of the paper is organized as follows. 
Section \ref{S2} introduces the notations and the necessary definitions.
Section \ref{S3} explains our quantum digital signature protocol. 
Section \ref{S4}
analyzes the security of the protocol.
Asymmetry, information-theoretical unforgeability, undeniability, and expandability are satisfied in the protocol. 
Section \ref{S5} analyzes the quantum state estimation given $n$ copies of the public key.
The last section gives a conclusion and discusses our future works.

\section{Preliminary}\label{S2}
\subsection{Notations}
We denote the quantum state with the Dirac symbol, e.g. $|\varphi\rangle$. We use $\{\cdot\}$ to represent a set.
We denote the probability for the appearance of an event $X$ as $P(X)$, we use ${n \choose k}$ to represent a combinatorial
number. $\{\Pi_{x}\}$ is used to denote a positive operator valued measure(POVM) set.

\subsection{Definition for quantum digital signatures}
We use a standard definition introduced in 
the preceding studies for quantum digital signatures. 
\begin{defn}
\name{(Quantum digital signatures \cite[Definition 4.2]{morimae2022},
\cite[Definition 4.1]{MY22}, \cite[Definition 9]{CM24})}
A quantum digital signature is
a set of quantum polynomial time algorithms (SKGen, PKGen, Sign, Ver) such that:

\textbullet SKGen $\left(1^\lambda\right) \rightarrow sk$  : It outputs a classical secret key $sk$ with the input of a security
 parameter $\lambda$.

\textbullet PKGen $(sk) \rightarrow pk$ : It outputs a quantum public key pk with the input of $sk$.

\textbullet Sign $(sk, m) \rightarrow s$: It outputs a classical signature $s$ with the inputs of $sk$ and a 
message $m$.

\textbullet $\operatorname{Ver}(pk, m, s) \rightarrow T / \perp$ : It outputs $T / \perp$ with 
the inputs of $pk, m$, and $s$.
\end{defn}

Notice that if there is no trusted third-party, a digital signature protocol should obey a 
basic assumption that the public keys are intactly transmitted.
As formulated in \cite{morimae2022,Morimae,MY22,CM24,KT23},
we assume that the adversary cannot modify the quantum public key
to be distributed to players.
This assumption is widely accepted in the community of cryptography.

\subsection{Definitions for security analysis}

As an alternative concept of one-way function, we introduce the following concept.

\begin{defn}
	\name{(Information-theoretic quantum one-way function)}
A function $f$ from classical information to quantum states
is called an information-theoretic quantum one-way function
when it is hard to identify the input of the function from
a limited number of copies of the output state of the function.
\end{defn}

	We define the quantum states $|0\rangle$ and $|1\rangle$ are the eigenstates of $\sigma_z$ corresponding to eigenvalues $1$ and $-1$ respectively,
	$|+\rangle$ and $|-\rangle$ are the eigenstates of $\sigma_x$ corresponding to eigenvalues $1$ and $-1$ respectively, 
	where $\sigma_z$ and $\sigma_x$ are Pauli operators.
Hence, in our protocol, we employ the function from 
an element of $(\{+1,-1\} \times \{\sigma_x,\sigma_z\})^l$ to 
$l$ corresponding eigenstates
as our information-theoretic quantum one-way function.

\begin{defn}
	\name{(Forging attack)}
Assume that a signature protocol generates a message $m=\{m_1,...,m_l\},m_a\in\{0,1\}$, and a signature $s=\{s^1,...,s^l\}$.
The following operation by an adversary is called a forging attack;
The adversary generates 
pairs $\{m'_1,s'^1\},\ldots ,\{m'_l,s'^l\}$, where $\{m'_1,...,m_l'\}=m'\neq m$
so that the verifier accepts the pair $\{m'_1,s'^1\}$ as a correct pair of 
a message and a signature.
\end{defn}

Since we assume that 
the adversary can collude with a limited number $k$ of players,
we define information-theoretic unforgeability as follows.

\begin{defn}
	\name{(Information-theoretic unforgeability)}
	We say that the signature protocol
	is $(k,\epsilon)$-information-theoretically unforgeable 
	when an adversary having $k$ copies of the public key succeeds forging attacks with at most probability of $\epsilon$.
\end{defn}

\begin{defn}
	\name{(Denying attack)}
The following operation by a signer is called a denying attack;
The signer makes the verifier to reject the signature sent by the signer.
\end{defn}

\begin{defn}
	\name{(Expandability)}
	When the length of the message in the protocol is $l$, 
	we denote the sum of the bits and qubits consumed in one turn
	(from key generation to verification for one message) by $C(l)$. 
	The protocol is expandable when
	the relation $C(l)\leq bl$ holds with a constant $b$ for a sufficiently large integer $l$.
\end{defn}

\begin{figure}[htbp]
	\centering
	\includegraphics[width=0.5\textwidth]{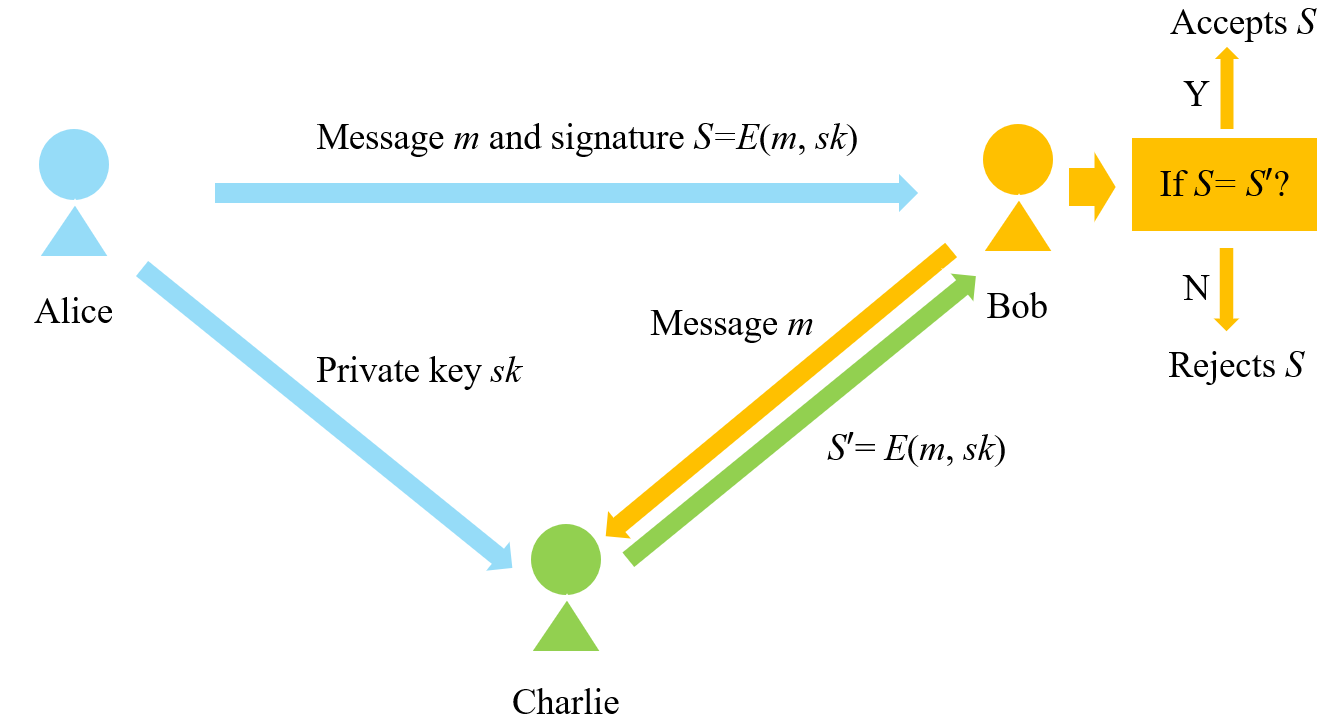}
	\caption{Quantum digital signature protocol with the first kind of trusted third-party. In this case, Alice works as the
	signer, Bob works as the verifier, and Charlie is the trusted third-party. In Key generation phase, Alice generates a 
	private key $sk$ and shares her private key $sk$ with Charlie. In Signing phase, Alice uses a function $E$ mapping a message
	$m$ and $sk$ to the signature $S$, and sends $m$ and $S$ to Bob. In Verification phase, Bob sends $m$ to Charlie, and
	Charlie uses the function $E$ to map $m$ and $sk$ to $S'$, and sends $S'$ to Bob. Finally, Bob compares $S$ with $S'$, 
	if $S=S'$, he accepts the signature, otherwise, he rejects the signature.}
	\label{F1}
\end{figure}

\begin{figure}[htbp]
	\centering
	\includegraphics[width=0.5\textwidth]{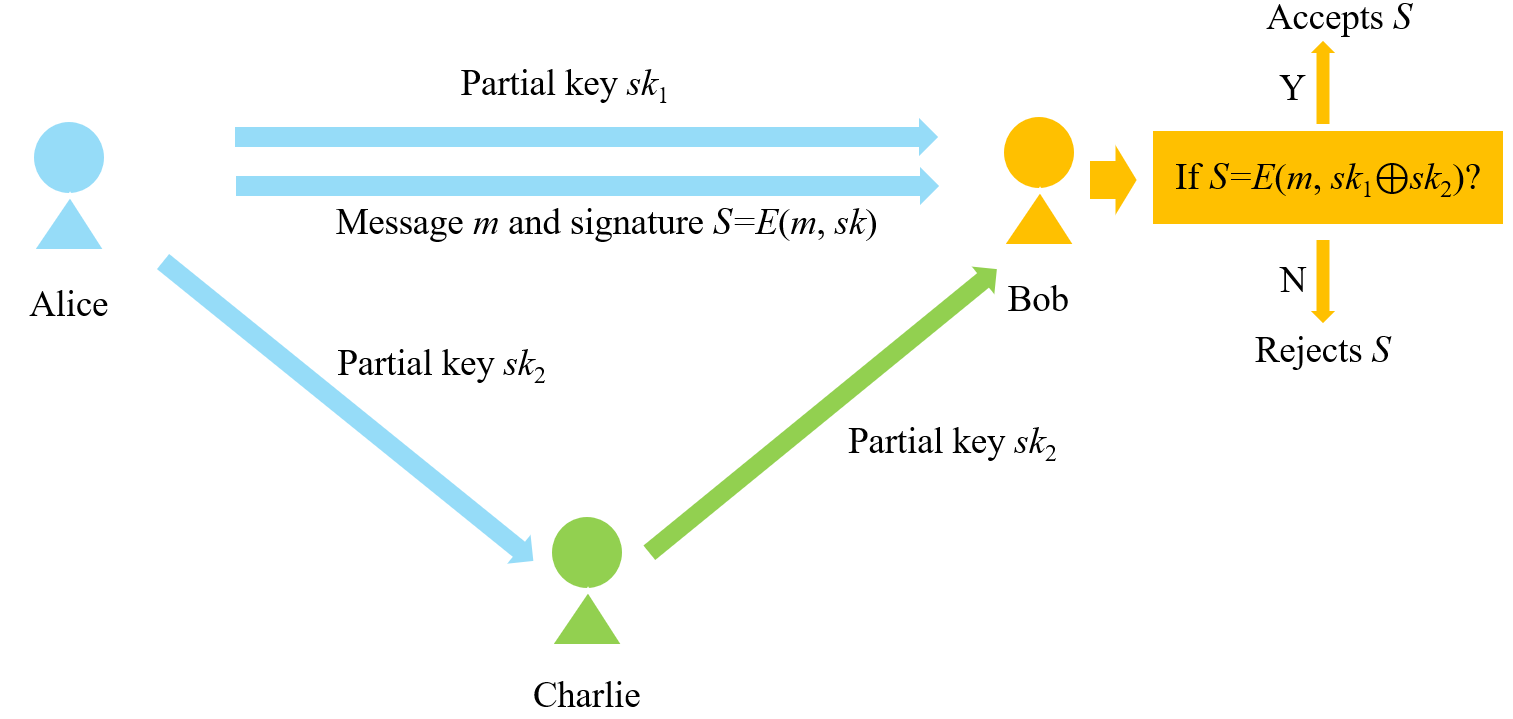}
	\caption{Quantum digital signature protocol with the second kind of trusted third-party. In this case, Alice works as the
	signer, Bob works as the verifier, and Charlie is the trusted third-party. In Key generation phase, Alice generates a 
	private key $sk$ and two partial keys $sk_1$ and $sk_2$ such that $sk=sk_1\oplus sk_2$. Then, Alice sends the partial key 
	$sk_1$ to Bob and the partial key $sk_2$ to Charlie. In Signing phase, Alice uses a function $E$ mapping a message
	$m$ and $sk$ to the signature $S$, and sends $m$ and $S$ to Bob. In Verification phase, Charlie sends the partial key
	$sk_2$ to Bob. At last, Bob judges if $S=E(m,sk_1\oplus sk_2)$, if the answer is yes, he accepts the signature, otherwise,
	he rejects the signature.}
	\label{F2}
\end{figure}

\section{Quantum digital signature protocol}\label{S3}
\begin{figure}[htbp]
	\centering
	\includegraphics[width=0.5\textwidth]{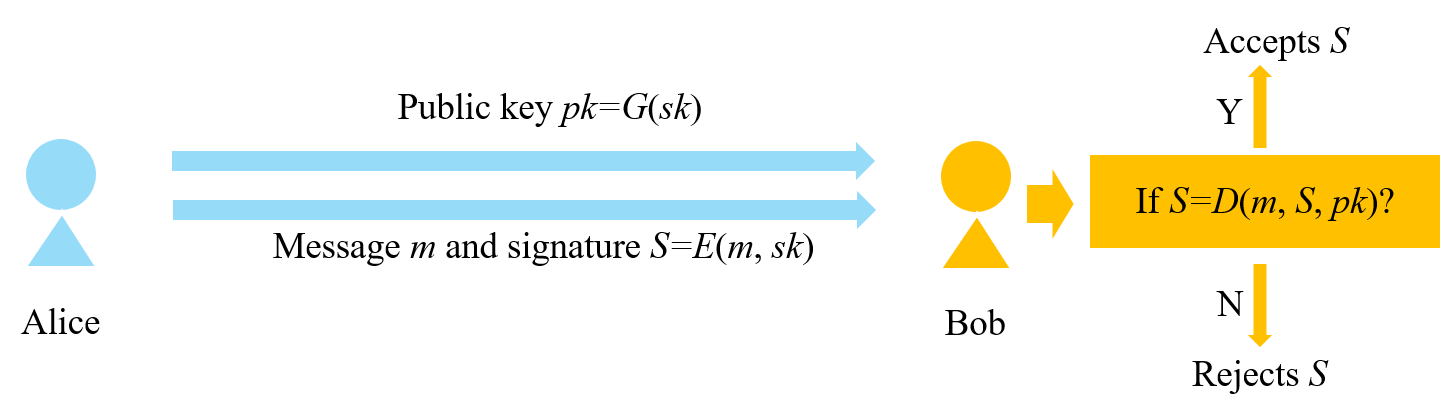}
	\caption{Quantum digital signature protocol without a trusted third-party. In this case, Alice works as the
	signer, Bob works as the verifier. In Key generation phase, Alice generates a private key $sk$ and use a function $G$
	to map $sk$ to a public key $pk$, then, she sends public key $pk$ to Bob. In Signing phase, Alice uses a function $E$ 
	mapping a message $m$ and $sk$ to the signature $S$, and sends $m$ and $S$ to Bob. In Verification phase, Bob uses a
	function $D$ to map $m$, $S$ and $pk$ to an outcome, then compares the value $S$ with the outcome $D(m,S,pk)$, if
	$S=D(m,S,pk)$, he accepts the signature, otherwise, he rejects the signature.}
	\label{F3}
\end{figure}

A trusted third-party in the Quantum digital signature protocol usually has two possible identities according to different protocols, the first kind of identity
shares the private key from the signer\cite{18}, as shown in Figure \ref{F1}, and the second kind of identity is the user supporting the secret sharing
with the verifier\cite{19}, as shown in Figure \ref{F2}. However, in our protocol, we remove the third-party while the protocol can complete
the main task, as shown in Figure \ref{F3}.

In our scenario, there are two types of players,
the signer, Alice, and $\mathcal{N}$ participants.
All participants receive public keys.
After the distribution of public keys, one of the participants, Bob, serves as the verifier to verify Alice's signature.
verifies Alice's sign. This participant is the verifier, Bob.
Alice and Bob negotiate a message length $l$ and a security parameter
$\lambda$($\lambda$ is a large enough positive integer). 

Our protocol
has three phases: Key generation phase, Signing phase, and Verification phase.
\subsection*{Key generation phase}
Alice performs $SKGen(1^\lambda)\rightarrow sk$ and $PKGen(sk)\rightarrow pk$.

\noindent Step 1: Alice chooses $l$ eigenvalue pairs $\left\{V P_1, \ldots, V P_l\right\}$ and $l$ basis pairs
$\{B P_1$, $\ldots$, $B P_l\}$, where $V P_a=\{ \{p_{a 10}, p_{a 11}\}$, $\ldots$,
$\{p_{a \lambda 0}, p_{a \lambda 1}\}\}$,
$BP_a=\{\{b_{a 10}, b_{a 11}\}$, $\ldots$, $\{b_{a \lambda 0}, b_{a \lambda 1}\}\}$, 
$p_{ijk}$ is
uniformly randomly chosen from the set $\{+1,-1\}$ and $b_{ijk}$ is uniformly randomly chosen from the set $\{\sigma_x,\sigma_z\}$, as the
private key.
That is, the length of the private key is $4l\lambda$ bits.
Here $sk=\{\left\{V P_1, \ldots, V P_l\right\},\{B P_1$, $\ldots$, $B P_l\}\}$ is the outputs of $SKGen(1^\lambda)$.

\noindent Step 2: 
Alice calculates $l$ eigenstate pairs $pk=\left\{S P_1, \ldots, S P_l\right\}$ 
as the public key, i.e., the outputs of $PKGen(sk)$,
where
$$S P_a=\left\{\left\{\left|e_{a 10}\right\rangle,\left|e_{a 11}\right\rangle\right\}, \ldots,\left\{\left|e_{a \lambda 0}
\right\rangle,\left|e_{a \lambda 1}\right\rangle\right\}\right\}$$
and $\left|e_{ijk}\right\rangle$ is the eigenstate for Pauli $b_{ijk}$ with the eigenvalue $p_{ijk}$. 

\noindent Step 3: 
Using $sk$, Alice prepares
the qubits $\{{\cal H}_{ijk}\}_{i=1,\ldots, l,j=1,\ldots, \lambda, k=0,1}$
in the eigenstates $\{\left|e_{ijk}\right\rangle\}
_{i=1,\ldots, l,j=1,\ldots, \lambda, k=0,1}$, and sends them to every participant.
That is, Alice repeats this procedure for $\mathcal{N}$ times.

\subsection*{Signing phase}
Alice performs $Sign(sk,m)\rightarrow s$.

\noindent Step 1: Alice prepares the message $m=\left\{m_1, \ldots, m_l\right\}, m_a \in\{0,1\}$.

\noindent Step 2: Alice chooses $s^1, \ldots, s^l$, where
$s^a=\left\{\left\{p_{a 1 m_a}, b_{a 1 m_a}\right\}, \ldots,\left\{p_{a \lambda m_a}, b_{a \lambda m_a}\right\}\right\}$ as
the signature.
Here $s=\{s^1, \ldots, s^l\}$ is the outputs of $Sign(sk,m)$.

\noindent Step 3: Alice sends the message-signature pairs $\left\{m_1, s^1\right\}, \ldots,\left\{m_l, s^l\right\}$ to Bob.

\subsection*{Verification phase}
Bob performs $Ver(pk,m,s)\rightarrow T/\perp$.

\noindent Step 1: Bob measures 
the qubits $\{{\cal H}_{acm_a}\}_{a=1,\ldots, l,c=1,\ldots, \lambda}$
with bases $\{b_{a c m_a}\}_{a=1,\ldots, l,c=1,\ldots, \lambda}$, respectively. If the measurement result of 
each qubit ${\cal H}_{acm_a}$ is the eigenstate corresponding to the eigenvalue $p_{a c m_a}$, he accepts the 
signature, else he rejects the signature.
Here the acceptance and rejection of the signature correspond to the outputs $T$ and $\perp$ of $Ver(pk,m,s)$, respectively.

\section{Security analysis}\label{S4}
In this section, we demonstrate that our protocol satisfies 
three security requirements:  
unforgeability, undeniability and expandability.

\subsection*{Unforgeability}
Our security analysis on unforgeability
is based on the following asymmetry.
The secret keys used in the protocol consist of the private key and the public key.
In the protocol, the private key is $l$ eigenvalue pairs $\left\{V P_1, \ldots, V P_l\right\}$ and $l$ basis pairs
$\left\{B P_1, \ldots, B P_l\right\}$, the public key is $l$ eigenstate pairs $\left\{S P_1, \ldots, S P_l\right\}$.
We note that one can directly prepare the quantum states $SP_a$ if he knows the eigenvalue pairs $VP_a$ and the basis pairs
$BP_a$. 
However, it is not so easy to know the private key $VP_a$ and $BP_a$ from $SP_a$ because he cannot know the measurement basis for each eigenstate. 
Therefore, the map from the private key $VP_a$ and $BP_a$ to the public key $SP_a$ works as an information-theoretic quantum one-way function.
The security of our protocol is based on this function.
However, 
when an adversary, Jack, colludes with 
too many participants holding public keys,
he can get the information of the corresponding private key.
Our security analysis depends on 
the number $n$ of copies of public keys that Jack can access as follows.

\begin{thm}
	The protocol presented in Section \ref{S3} is $\left(0,\left(\frac{1}{2}\right)^\lambda\right)$,
	and $\left(n,F(n)^\lambda\right)$
	-information-theoretically unforgeable with $n\geq 1$, where $F(n):=\frac{1}{2}+\frac{1}{2^{n+1}} 
\sum_{j \in \mathbb{Z}_4} 
\alpha_{j,n}^{1/2}
\alpha_{j+1,n}^{1/2}$
and
	$\alpha_{j,n}:=
	\sum_{k: k=j \hbox{ \rm mod} 4}
	{n \choose k}$ for $j=0,1,2,3$.
\end{thm}

\begin{figure}[htbp]
	\centering
	\includegraphics[width=0.5\textwidth]{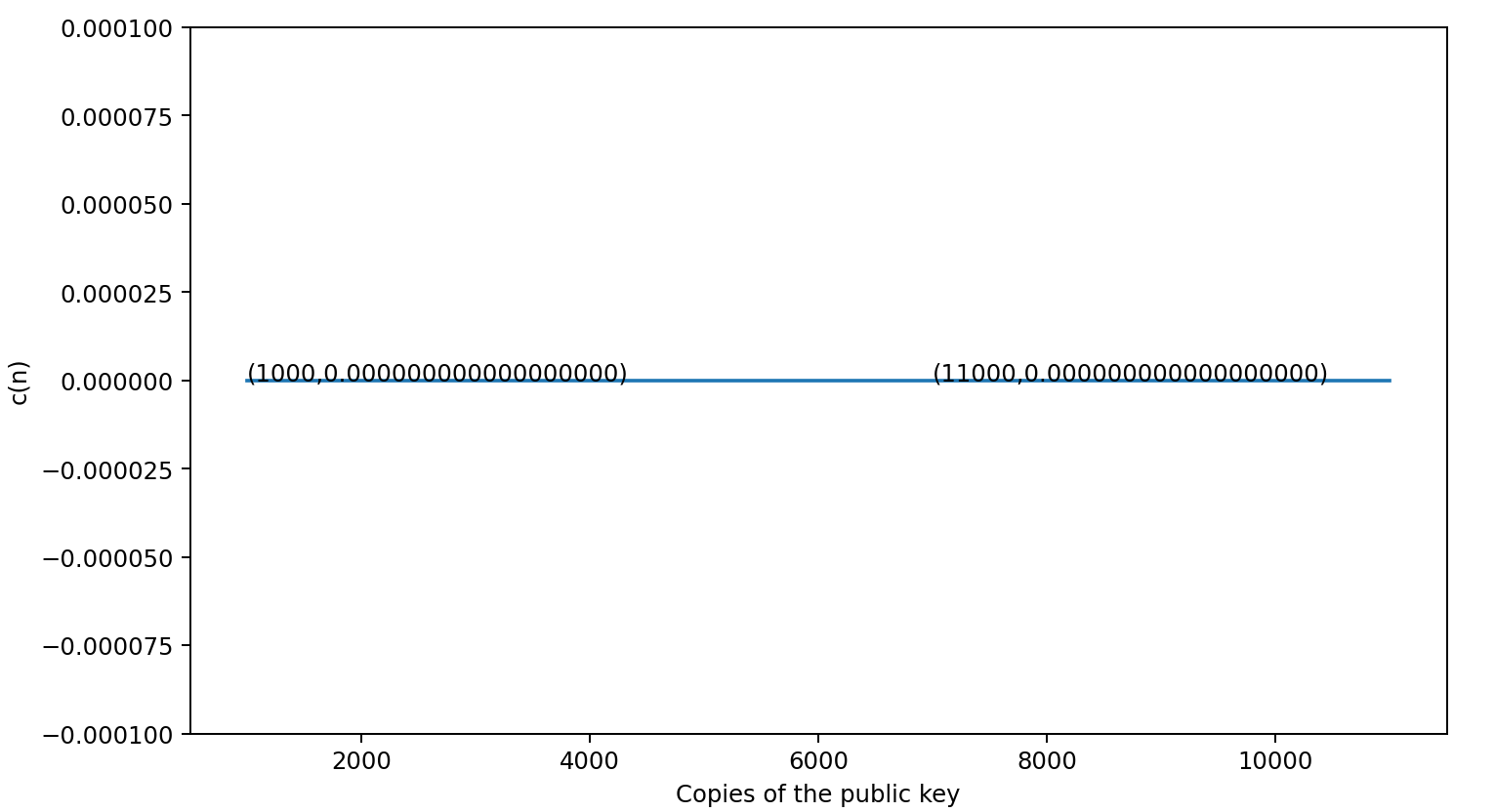}
	\caption{The numerical calculation for the function $c(n)=\frac{\log (1-F(n))}{-(n+1)}-1$. 
	The x-axis presents $n$ and the y-axis presents $c(n)$. 
	We note that $F(n)$ is smaller than $1$ and goes to $1$ as the copies of the public key increase
	. It can be found that the behavior of the function $\log (1-F(n))$ is approximated to $-c(n+1)$ with $c=1.00000$.
	}
	\label{F4}
\end{figure}

To ensure the unforgeability, 
we discuss how to choose an appropriate $\lambda$. 
First of all, we do the numerical calculation for the function $\frac{\log (1-F(n))}{-(n+1)}$, as shown in Figure \ref{F4}.
This numerical calculation suggests the approximation
$\log (1-F(n))=-c(n+1)$, where $c:=1.00000$.
That is, $F(n)$ is approximated to $1- 2^{-c(n+1)}$.

\begin{figure}[htbp]
	\centering
	\includegraphics[width=0.5\textwidth]{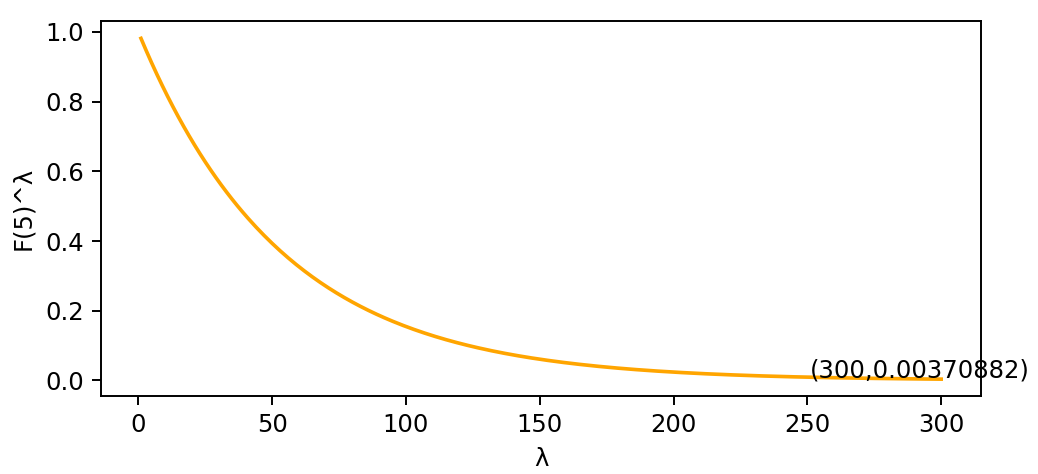}
	\caption{The numerical calculation for the function $F(5)^{\lambda}$. We note that if we let $\lambda=300$,
	the signature forged by an adversary can be accepted with at most the probability of $0.0370882$.
	}
	\label{F5}
\end{figure}

Fig \ref{F4} shows that the security measure $F(n)^\lambda$ is approximated to $(1- 2^{-c(n+1)})^{\lambda}$.
When the number $n$ of copies is large, 
the number $\lambda$ needs to be a quite large number, which is not practical.
However, when the number $n$ of copies is not so large, i.e., 
is at most $5$,
we can realize a reasonable security parameter 
$F(5)^{\lambda}$ with reasonable size of $\lambda$
as Figure \ref{F5}.
Therefore, 
we can say that our proposed protocol is secure against 
forging attack
when an adversary, Jack, colludes with a limited number of 
participant holding public keys.

\begin{proof}
We consider the case for $k=0$ first.
We assume that there is an adversary Jack, who wants to generate pairs 
$\left\{m'_1, s'^1\right\}, \ldots,\left\{m'_l, s'^l\right\}$ to replace pairs 
$\left\{m_1, s^1\right\}, \ldots,\left\{m_l, s^l\right\}$.
Given $a=1, \ldots, l$,
let $P(S)$ be the probability for Bob to accept the signature forged by Jack,
$P(S_{a,c})$ be the probability that the pair $\left\{p_{a c m_a}', b_{a c m_a}'\right\}$ is accepted
under the eigenstate $\left|e_{a c m_a}\right\rangle$,
and $P(S_a)$ be the probability that the signature $s'^a
=\left\{\left\{p_{a 1 m_a}', b_{a 1 m_a}'\right\}, \ldots,
\left\{p_{a \lambda m_a}', b_{a \lambda m_a}'\right\}\right\}
$ is accepted by Bob.

For the case $m'_a\neq m_a$, whatever the choice of the pair $\left\{p_{a c m_a}^{\prime}, b_{a c m_a}^{\prime}\right\}$ is for $c=1,\ldots, \lambda$,
$P(S_{a,c})=1/2$. For example, if he chooses the pair $\left\{1,Z\right\}$, we have 
$P(S_{a,c}\mid\left|e_{a c m_a}\right\rangle=\left|0\right\rangle)=1$,
$P(S_{a,c}\mid\left|e_{a c m_a}\right\rangle=\left|1\right\rangle)=0$,
$P(S_{a,c}\mid\left|e_{a c m_a}\right\rangle=\left|+\right\rangle)=1/2$, and
$P(S_{a,c}\mid\left|e_{a c m_a}\right\rangle=\left|-\right\rangle)=1/2$.
Since the probability distribution for
$\left|e_{a c m_a}\right\rangle$ is $\left\{1/4,1/4,1/4,1/4\right\}$ corresponding to $\left|0\right\rangle,\left|1\right\rangle,
\left|+\right\rangle,\left|-\right\rangle$ respectively, we have $P(S_{a,c})=1/2$. Similarly, we have the same result for pairs
$\left\{-1,Z\right\},\left\{1,X\right\},\left\{-1,X\right\}$.
	Since every eigenstate is generated independently, we can calculate $P(S_a)$ as
\begin{equation}
	P\left(S_a\right)=
	\prod_{c=1}^{\lambda}P\left(S_{a,c}\right)=\left(\frac{1}{2}\right)^\lambda.
\end{equation}

For the case $m'_{a_1}\neq m_{a_1},\ldots,m'_{a_K}\neq m_{a_K}
(a_1< a_2< \cdots< a_K)$ with a positive 
integer $K\leq l$, we have
\begin{equation}
	P\left(S_{a_1}, \ldots, S_{a_K}\right)=P\left(S_{a_1}\right) P\left(S_{a_2}\right) \cdots P\left(S_{a_K}\right)=
	\left(\frac{1}{2}\right)^{K \lambda} \leq\left(\frac{1}{2}\right)^\lambda.
\end{equation}

Then, we define a variable $X_K=(a_1,\ldots,a_K)$ corresponding to the case 
$m'_{a_1}\neq m_{a_1},\ldots,m'_{a_K}\neq m_{a_K}(a_1\neq a_2\neq \cdots\neq a_K)$, the probability $P(S)$ can be 
calculated as follows.
\begin{align}
	P(S)&=\sum_{i=1}^l\sum_{X_i} P(S_{a_1},\ldots,S_{a_i}) P(X_i=(a_1,\ldots,a_i))\notag\\
	&\leq P(S_{a_1}) \sum_{i=1}^l\sum_{X_i} P(X_i=(a_1,\ldots,a_i))=\left(\frac{1}{2}\right)^\lambda.
\end{align}

Next, we assume that Jack has $n(n\geq 1)$ copies of the public key $\left\{S P_1, \ldots, S P_l\right\}$.
In this case, $P(S_{a,c})$ is the probability that
Jack correctly identifies the true state among the four states
$\left|0\right\rangle,\left|1\right\rangle,
\left|+\right\rangle,\left|-\right\rangle$ 
from $n$ copies of the true state.
The maximum value of this probability is given in 
Theorem \ref{GHT}, which is discussed in Section \ref{S5}.
Hence, applying the Theorem \ref{GHT}, we have
$1-P\left(S_{a,c}\right)=\frac{1}{2}-
	\frac{1}{2^{n+1}} 
	\sum_{j \in \mathbb{Z}_4} 
	\alpha_{j,n}^{1/2}
	\alpha_{j+1,n}^{1/2}$, which implies
\begin{equation}
	P(S)\leq P(S_{a})=
	\prod_{c=1}^{\lambda}P\left(S_{a,c}\right)=\left(\frac{1}{2}+
	\frac{1}{2^{n+1}} 
	\sum_{j \in \mathbb{Z}_4} 
	\alpha_{j,n}^{1/2}
	\alpha_{j+1,n}^{1/2}\right)^\lambda.
\end{equation}
\end{proof}

\subsection*{Undeniability}
The undeniability of our protocol is automatically satisfied since the communications between Alice and Bob only exist
in Key generation phase and Signing phase. Once Alice sends the public key, message, and signature to Bob, Bob can always
accept the signature.

\subsection*{Expandability}
In the protocol presented in Section \ref{S3},
we denote the bits consumed for the private key and the signature as $C_1(l)$ and $C_2(l)$ respectively, and
the qubits consumed for the public key as $C_3(l)$. 
Now we have $C_1(l)=4\lambda l$, $C_2(l)=2 \lambda l$, $C_3(l)=2\lambda l$. Hence, we have
\begin{align}
	C(l)=C_1(l)+C_2(l)+C_3(l)=8\lambda l.
\end{align}
Since $\lambda$ is a constant, according to Definition 4, the expandability is also satisfied.

\section{State identification}\label{S5}
In this section, we assume that the adversary Jack has $n$ copies of a quantum state uniformly
distributed on $|1\rangle,|0\rangle,|+\rangle,|-\rangle$ and discuss the minimum identification error probability for
the quantum state.
\subsection{Minimum error}
The security analysis of our protocol
requires the maximum value of the state identification
of the following type of problem setting.
We choose the eigenvectors  
$|e_0\rangle := \frac{1}{\sqrt{2}}(|0\rangle +i|1\rangle)$
and $|e_1\rangle := \frac{1}{\sqrt{2}}(|0\rangle -i|1\rangle)$
of $Y$.
Then, we have
\begin{align}
|0\rangle&= \frac{1}{\sqrt{2}}( |e_0\rangle+|e_1\rangle) ,\quad
\frac{1+i}{\sqrt{2}}|+\rangle= \frac{1}{\sqrt{2}}( |e_0\rangle+i|e_1\rangle) \\
i |1\rangle&= \frac{1}{\sqrt{2}}( |e_0\rangle-|e_1\rangle) ,\quad
\frac{1-i}{\sqrt{2}}|-\rangle= \frac{1}{\sqrt{2}}( |e_0\rangle-i|e_1\rangle) .
\end{align}
Therefore, our possible states form
the state family $\{ | f_x\rangle\}_{x \in \mathbb{Z}_4}$, where
$| f_x\rangle:= \frac{1}{\sqrt{2}}( |e_0\rangle+e^{x \pi  i/2}
|e_1\rangle)$.
We consider estimation problem under this state family.
Given the true parameter $x$ and our estimate $\hat{x}$, 
we employ the following error function:
\begin{align}
R(x,\hat{x}):=\frac{1-\cos (x-\hat{x})\frac{\pi}{2}}{2}
=\frac{1}{2}-\frac{1}{4}(e^{\frac{(x-\hat{x}) \pi i}{2}} +e^{-\frac{(x-\hat{x}) \pi i}{2}} )
\end{align}
That is, when the true parameter is $x$, 
and the estimate by Jack is $\hat{x}$,
the probability that the information sent by Jack coincides with
Bob's measurement outcome is $1-R(x,\hat{x})$.
Therefore, to consider Jack's optimal attack, 
we need to minimize the average of $R(x,\hat{x})$.

In the one-copy case, 
we need to consider the following minimization 
\begin{align}
\min_{ \{\Pi_{\hat{x}}\} }
\sum_{x=0}^3 \frac{1}{4}
\sum_{\hat{x}=0}^3 
R(x,\hat{x}) \Tr \Pi_{\hat{x}} | f_x\rangle \langle f_x|,
\end{align}
where $\{\Pi_{\hat{x}}\}$ is a POVM over one-qubit system.
In the $n$-copy case, 
we need to consider the following minimization 
\begin{align}
\min_{ \{\Pi_{\hat{x}}\} }
\sum_{x=0}^3 \frac{1}{4}
\sum_{\hat{x}=0}^3 
R(x,\hat{x}) \Tr \Pi_{\hat{x}} (| f_x\rangle \langle f_x|)^{\otimes n},
\end{align}
where $\{\Pi_{\hat{x}}\}$ is a POVM over $n$-qubit system.
For $j \in \mathbb{Z}_4$, We define 
\begin{align}
\alpha_{j,n}:=
\sum_{k: k=j \hbox{ \rm mod} 4}
{n \choose k}.
\end{align}
Then, we obtain 
the minimum identification error probability as follows.
\begin{thm}\label{GHT}
We have the following relation
 \begin{align}
\min_{ \{\Pi_{\hat{x}}\} }
\sum_{x=0}^3 \frac{1}{4}
\sum_{\hat{x}=0}^3 
R(x,\hat{x}) \Tr \Pi_{\hat{x}} (| f_x\rangle \langle f_x|)^{\otimes n}
=\frac{1}{2}-
\frac{1}{2^{n+1}} 
\sum_{j \in \mathbb{Z}_4} 
\alpha_{j,n}^{1/2}
\alpha_{j+1,n}^{1/2}.\label{BGU}
\end{align}
\end{thm}

\subsection{Proof of Theorem \ref{GHT}}
To prove Theorem \ref{GHT}, we employ 
the covariant state estimation theory \cite[Chapter 4]{Holevo},\cite[Chapter 4]{Group2}.

We consider the representation $V_x $ of the group $ \mathbb{Z}_4$ as
\begin{align}
V_x:= |e_0\rangle\langle e_0| +e^{x \pi i/2}|e_1\rangle \langle e_1|.
\end{align}
Then, the state family $\{ 
(| f_x\rangle \langle f_x|)^{\otimes n}
\}_{x \in \mathbb{Z}_4}$
and the error function $R(x,\hat{x})$ satisfies
the group covariance with respect to the representation $V_x^{\otimes n}$.
The vector $| f_x\rangle^{\otimes n}$
belongs to the $n$-th symmetric subspace ${\cal H}_n$, which has the dimension $n+1$
and is spanned by
$\{ | g_{k,n}\rangle \}_{k=0}^n$, where
\begin{align}
| g_{k,n}\rangle:= {n \choose k}^{-1/2} (  
\underbrace{|e_0\rangle \cdots |e_0\rangle}_{k} 
\underbrace{|e_1\rangle \cdots |e_1\rangle}_{n-k}  + permutation).
\end{align}
The state $| f_x\rangle ^{\otimes n}$ is written as
\begin{align}
&| f_x\rangle ^{\otimes n}
= \sum_{k=0}^n 
\frac{1}{2^{n/2}} {n \choose k}^{1/2} e^{ k x \pi i/2 }| g_{k,n}\rangle \notag\\
=&
\sum_{j \in \mathbb{Z}_4 }
\sum_{k: k=j \hbox{ \rm mod} 4}
\frac{1}{2^{n/2}} {n \choose k}^{1/2} e^{ k x \pi i/2 }| g_{k,n}\rangle \notag\\
=&
\sum_{j \in \mathbb{Z}_4 }
\frac{1}{2^{n/2}} \alpha_{j,k}^{1/2} e^{ j x \pi i/2 }|\beta_{j,n}\rangle,
\end{align}
where the normalized vector $|\beta_{j,n}\rangle$ is defined as
\begin{align}
|\beta_{j,n}\rangle:=
\alpha_{j,k}^{-1/2}
\sum_{k: k=j \hbox{ \rm mod} 4}
 {n \choose k}^{1/2} | g_{k,n}\rangle .
\end{align}
Hence, the state 
$| f_x\rangle ^{\otimes n}$ belongs to the subspace 
${\cal K}_{n}$
spanned
by $\{|\beta_{j,n}\rangle\}_{j \in \mathbb{Z}_4}$.

The representation on ${\cal K}_n$
is given as
\begin{align}
U^{(n)}_{x}:=
\sum_{j \in \mathbb{Z}_4 }
e^{ j x \pi i/2 } |\beta_{j,n}\rangle \langle \beta_{j,n}| .
\end{align}
Hence, the optimal estimator is given as a covariant measurement over 
${\cal H}_n$.
That is, the optimal POVM has the form 
$\{ U^{(n)}_{\hat{x}} T U^{(n)}_{-\hat{x}}\}_{\hat{x}=0}^3$,
where $T \ge 0$ and 
$ \langle \beta_{j,n}|T |\beta_{j,n}\rangle =1/4$ for $j \in \mathbb{Z}_4$.
Then, to characterize the off-diagonal part of $T$, 
we introduce 
\begin{align}
a_{j,j'}:=4 \langle \beta_{j,n}|T |\beta_{j',n}\rangle,
\end{align}
which satisfies $|a_{j,j'}|\le 1$.
 
In the following, we consider 
$(e^{\frac{(x-\hat{x}) \pi i}{2}} +e^{-\frac{(x-\hat{x}) \pi i}{2}} )$
instead of 
$R(x,\hat{x})
=\frac{1}{2}-\frac{1}{4}
(e^{\frac{(x-\hat{x}) \pi i}{2}} +e^{-\frac{(x-\hat{x}) \pi i}{2}} )$.
  That is, we maximize 
$\sum_{x=0}^3 \frac{1}{4}
\sum_{\hat{x}=0}^3 
(e^{\frac{(x-\hat{x}) \pi i}{2}} +e^{-\frac{(x-\hat{x}) \pi i}{2}} )
 \langle f_x|^{\otimes n}\Pi_{\hat{x}} | f_x\rangle ^{\otimes n}$.  
  
Using the relation
\begin{align}
\sum_{\hat{x}=0}^3 
(e^{-\frac{\hat{x} \pi i}{2}} +e^{\frac{\hat{x} \pi i}{2}} )
e^{ (j-j') \hat{x} \pi i/2 }
=
\left\{
\begin{array}{ll}
1 & \hbox{ when }  j-j'=1, 3 \hbox{ mod }4 \\
0 & \hbox{ otherwise},
\end{array}
\right.
\end{align}
we have
\begin{align}
&\frac{1}{4}\sum_{x=0}^3 
\sum_{\hat{x}=0}^3 
(e^{\frac{(x-\hat{x}) \pi i}{2}} +e^{-\frac{(x-\hat{x}) \pi i}{2}} )
 \langle f_x|^{\otimes n}\Pi_{\hat{x}} | f_x\rangle ^{\otimes n} 
 \notag\\
=&\frac{1}{4}
\sum_{\hat{x}=0}^3 
(-e^{\frac{\hat{x} \pi i}{2}} +e^{\frac{\hat{x} \pi i}{2}} )
 \langle f_0|^{\otimes n}\Pi_{\hat{x}} | f_0\rangle ^{\otimes n}\notag\\
=&
\frac{1}{4}\sum_{j,j' \in \mathbb{Z}_4}
\sum_{\hat{x}=0}^3 
(e^{-\frac{\hat{x} \pi i}{2}} +e^{\frac{\hat{x} \pi i}{2}} )
\langle f_0|^{\otimes n}
a_{j,j'}
U^{(n)}_{\hat{x}}
|\beta_{j,n}\rangle \langle \beta_{j',n}| 
U^{(n)}_{-\hat{x}}
 | f_0\rangle ^{\otimes n}
\notag\\
=&
\frac{1}{4}\frac{1}{2^{n}} 
\sum_{j,j' \in \mathbb{Z}_4}
\sum_{\hat{x}=0}^3 
(e^{-\frac{\hat{x} \pi i}{2}} +e^{\frac{\hat{x} \pi i}{2}} )
e^{ (j-j') \hat{x} \pi i/2 }
a_{j,j'}
\alpha_{j,n}^{1/2}
\alpha_{j',n}^{1/2}
\notag\\
=&
\frac{1}{4}\frac{1}{2^{n}} 
\sum_{j,j' \in \mathbb{Z}_4:j-j'=1,3}
a_{j,j'}
\alpha_{j,n}^{1/2}
\alpha_{j',n}^{1/2}
\notag\\
=&
\frac{1}{4}\frac{1}{2^{n}} 
\sum_{j,j' \in \mathbb{Z}_4:j-j'=1,3}
\frac{1}{2}(a_{j,j'}+a_{j',j})
\alpha_{j,n}^{1/2}
\alpha_{j',n}^{1/2}
\notag\\
\le &
\frac{1}{4}\frac{1}{2^{n}} 
\sum_{j,j' \in \mathbb{Z}_4:j-j'=1,3}
\alpha_{j,n}^{1/2}
\alpha_{j',n}^{1/2}
=\frac{1}{2^{n+1}} 
\sum_{j \in \mathbb{Z}_4}
\alpha_{j,n}^{1/2}
\alpha_{j+1,n}^{1/2}.
\end{align}
The equality holds when $ a_{j,j'}=1$.
Therefore, we conclude \eqref{BGU}.

\section{Conclusion and future works}
Quantum digital signature plays an important role in constructing advanced quantum information systems. A basic task
required in many quantum information systems is the authentication of the identity of a user. 
A trusted third-party is necessary for many known quantum digital signature schemes \cite{15,16,17,18,19,20}. 
To break through the bottleneck of
the security, we propose a qubit-based quantum digital signature protocol without the third-party. On this basis, our protocol satisfies unforgeability
when there are not so many copies of the public keys in the adversary's hand.
It also satisfies other significant security requirements: undeniability and expandability.
We conclude the security comparison with preceding studies in Table \ref{T1}.
In addition, 
our protocol can be implemented only with single qubit physical operations, which does not require any entangling operation.
It should be mentioned
that our protocol can be further improved because there is room for reduction of $C(l)$. 
The security analysis
for the combination of quantum digital signatures with other quantum information systems needs to be developed. Hence, in
the future, our work focus on how to solve the above problems.
In addition, 
although our protocol is given as a simple combination of qubit technologies,
it requires quantum memories.
Implementation of quantum memories is a key technology for our protocol.

\section*{Acknowledgement}
W.W. was supported in part by 
JST SPRING, Grant Number JPMJSP2125
and the “THERS Make New Standards Program for the Next Generation Researchers.” 
M.H. was supported in part by the National Natural
Science Foundation of China under Grant 62171212
and
the General R\&D Projects of 
1+1+1 CUHK-CUHK(SZ)-GDST Joint Collaboration Fund 
(Grant No. GRDP2025-022).

\section*{References}
\bibliographystyle{iopart-num}
\bibliography{references}
\end{document}